\documentclass{webofc}

\usepackage[varg]{txfonts}   
\usepackage{hyperref}
\usepackage{url}
\hypersetup{colorlinks=true,citecolor=blue,urlcolor=blue,linkcolor=blue}
\newcommand{\cref}[1]{(\ref{#1})}

\newcommand{\ev}[1]{\langle #1 \rangle}

\begin{document}
\title{Equation of state, QCD phase diagram: predictions from lattice QCD}
%
%

\author{\firstname{Bastian B.} \lastname{Brandt}\inst{1}\fnsep\thanks{\email{brandt@physik.uni-bielefeld.de}}}

\institute{Fakult\"at f\"ur Physik, Universit\"at Bielefeld, 33615 Bielefeld, Germany}

\abstract{
I review recent results on phase structure and equation of state of strong interaction matter from lattice QCD. Particular emphasis is given to the axes where direct simulations are possible and results are obtained with sufficient control over systematic effects. I also discuss the status of approaching the region of non-zero baryochemical potentials using indirect methods.
}
\maketitle
\section{Introduction}
\label{intro}

Ultrarelativistic collisions of heavy ions (HICs) allow to investigate strong interaction matter under extreme conditions, such as high temperatures and/or large densities. Ab initio theoretical studies play a pivotal role for being able to interpret the experimental results. At the temperatures and densities achieved, quantum chromodynamics (QCD), the theory of the strong interactions, remains strongly coupled so that non-perturbative methods are needed to obtain reliable results. The two available non-perturbative ab initio methods are functional methods and numerical simulations of lattice QCD. While full QCD functional results with increased control over systematic effects start to become available, lattice QCD remains the only method which can be improved systematically. Results at vanishing chemical potentials can be obtained directly using Monte-Carlo methods, but the region of non-zero chemical potentials is plagued by the infamous sign problem and most regions can only be accessed indirectly.
This includes the phase diagram in the temperature-baryon chemical potential plane, $(T,\mu_B)$, see figure~\ref{fig-1}, and the detection of a possible critical endpoint (CEP).
I review recent lattice results for the QCD phase structure and the equation of state (EoS), the main input for a hydrodynamical description of the fireball created in the collisions.

\section{Predictions from lattice QCD}
\label{sec-1}

\subsection{Relevant parameter space}
\label{sec-1.1}

\begin{figure}
\centering
\sidecaption
\includegraphics[width=5.8cm,clip]{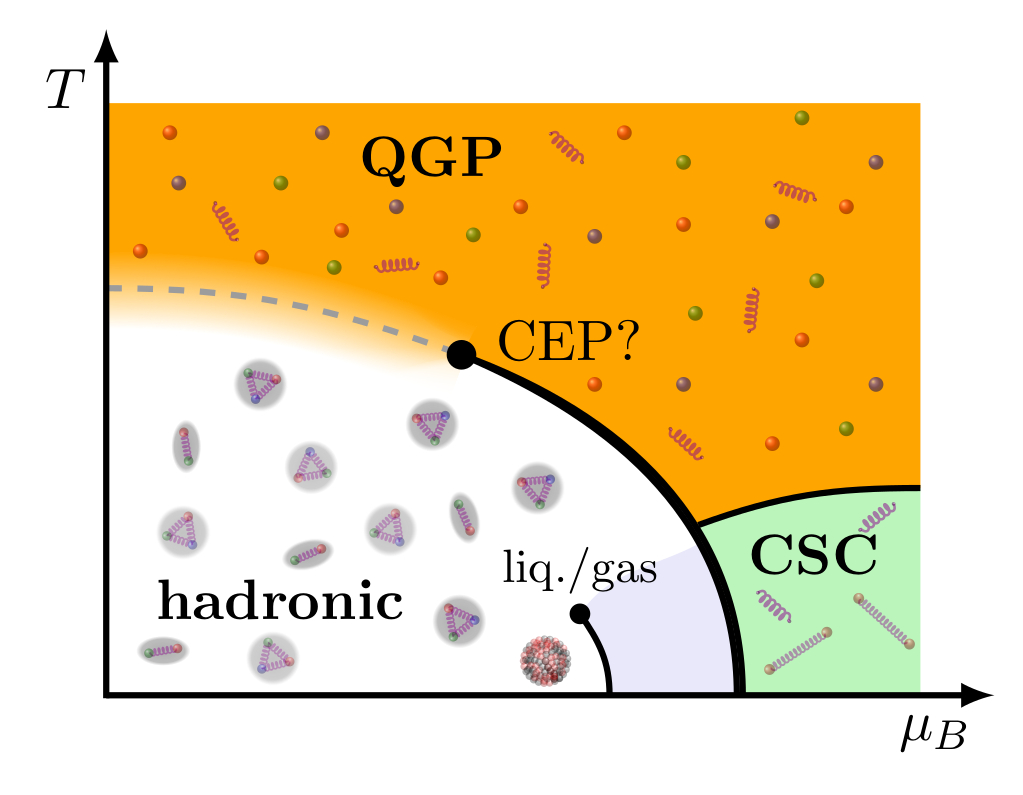}
\caption{Conjectured phase diagram in the plane of temperature $T$ and non-zero baryon chemical potential $\mu_B$, including the hadronic phase at small $T$ and $\mu_B$, the quark-gluon plasma (QGP) phase and possible phases with color superconductivity (CSC) at large $\mu_B$. The sketch also includes the possible critical endpoint (CEP) where the crossover at $\mu_B=0$ turns first order and the liquid/gas phase transition at small temperatures.}
\label{fig-1}       
\end{figure}

Of direct relevance for the QCD phase structure are the contributions of the three lightest quarks ($u$, $d$ and $s$), whereas heavier quarks only influence the EoS at larger temperatures $T$. In the grand canonical ensemble the relevant parameters are the quark chemical potentials, $\mu_q$, which can be conveniently expressed in the ``physical'' basis,
\begin{equation}
\mu_u=\frac{\mu_B}{3}+\frac{2\mu_Q}{3} \,, \qquad \mu_d= \frac{\mu_B}{3}-\frac{\mu_Q}{3} \quad\text{and}\quad \mu_s= \frac{\mu_B}{3}-\frac{\mu_Q}{3}-\mu_S \,,
\end{equation}
via baryon ($B$), charge ($Q$) and strangeness ($S$) chemical potentials. Most relevant for HIC physics is $\mu_B$, for which the conjectured phase diagram is shown in figure~\ref{fig-1}. In HICs, the average relative net densities $\ev{n_Q}/\ev{n_B}$ and $\ev{n_S}/\ev{n_B}$ are dictated by the relative densities of the respective ions in the collision region. For instance for Pb-Pb collisions $ \ev{n_Q}/\ev{n_B}\approx0.4$ and $\ev{n_S}=0$, corresponding to trajectories with non-zero $\mu_Q$ and $\mu_S$ in parameter space. For theoretical studies it is often more convenient to use the ``isospin'' basis,
\begin{equation}
\label{eq:isospin-base}
\mu_u=\frac{\mu_\mathcal{B}}{3}+\mu_I \,, \qquad \mu_d= \frac{\mu_\mathcal{B}}{3}-\mu_I \quad\text{and}\quad \mu_s= \frac{\mu_\mathcal{B}}{3}-\mu_\mathcal{S} \,,
\end{equation}
where $\mu_Q$ has been traded for the isospin chemical potential $\mu_I$\footnote{Different definitions for $\mu_I$ exist in the literature. For the one of eq.~\cref{eq:isospin-base} charged pions have isospin $\pm2$.} and baryon $\mu_\mathcal{B}$ and strangeness $\mu_\mathcal{S}$ generically differ from the ones in the physical basis. Lattice simulations suffer from the sign problem when $\mu_\mathcal{B}\neq 0 \neq \mu_\mathcal{S} $ and are sign problem free at pure isospin asymmetry, $\mu_I\neq0,\,\mu_\mathcal{B}=\mu_\mathcal{S}=0$, with degenerate light quarks. A number of physical systems, such as off-central HICs, potentially also feature magnetic fields $\mathbf{B}$ on the order of the QCD scale, which strongly influence the physical properties of the theory. The extension of the phase diagram in the direction of non-zero $\mu_I$ and external $\mathbf{B}$-fields is shown in figure~\ref{fig-2}.

\subsection{Direct results on the phase structure and the equation of state}
\label{sec-1.2}

The temperature axis has been investigated thoroughly in the past three decades from lattice QCD. The transition is found to be a crossover~\cite{Aoki:2006we} at a pseudocritical temperature $T_{pc}\sim158$~MeV~\cite{HotQCD:2018pds,Borsanyi:2020fev}. The EoS has been computed with high accuracy up to $T\lesssim2$ GeV~\cite{Borsanyi:2013bia,HotQCD:2014kol,Bazavov:2017dsy} and has recently been extended using novel step scaling methods up to $T\lesssim165$~GeV~\cite{Bresciani:2025vxw}.

\begin{figure}[t]
\centering
\includegraphics[width=5.5cm,clip]{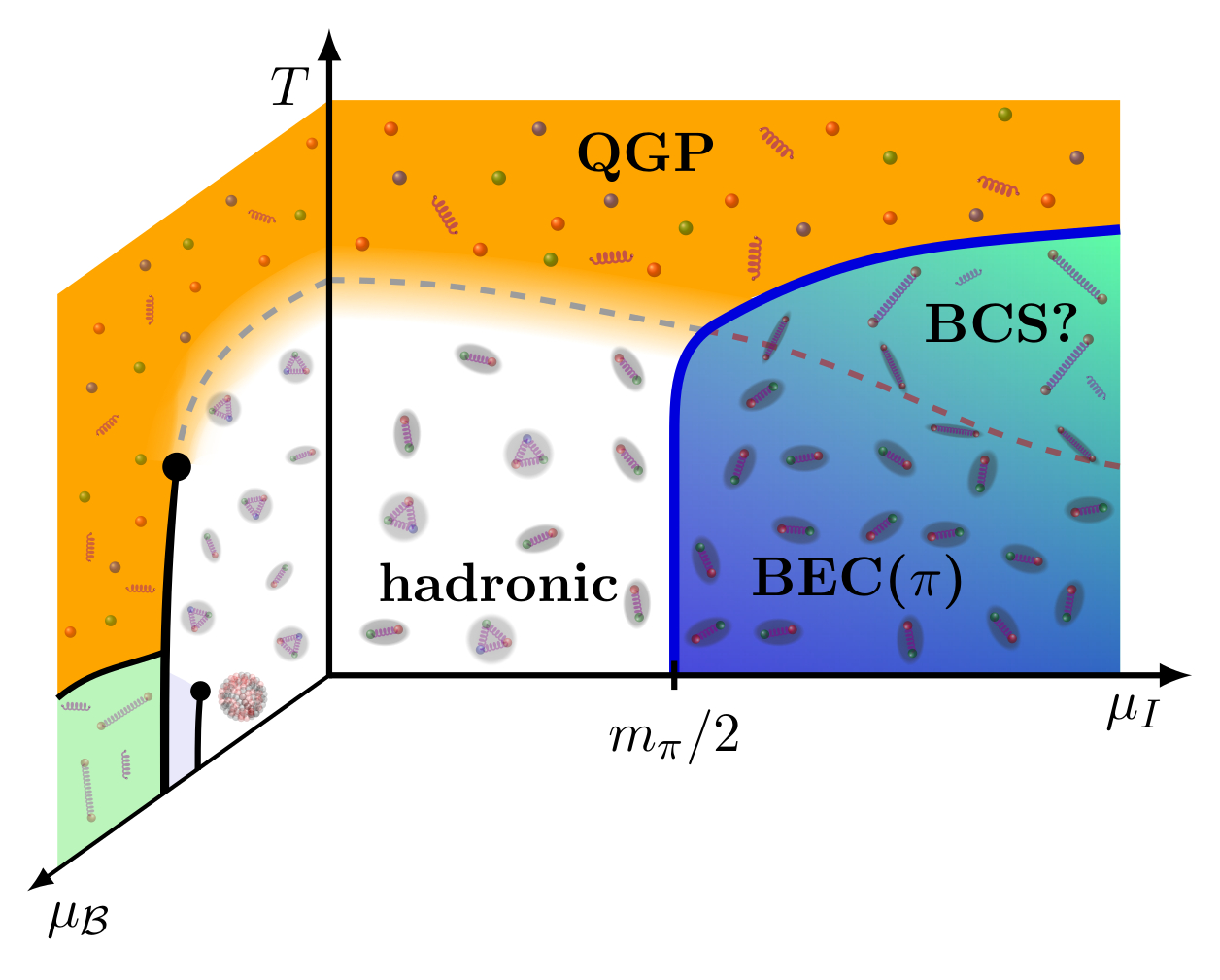} \hspace*{5mm}
\includegraphics[width=5.5cm,clip]{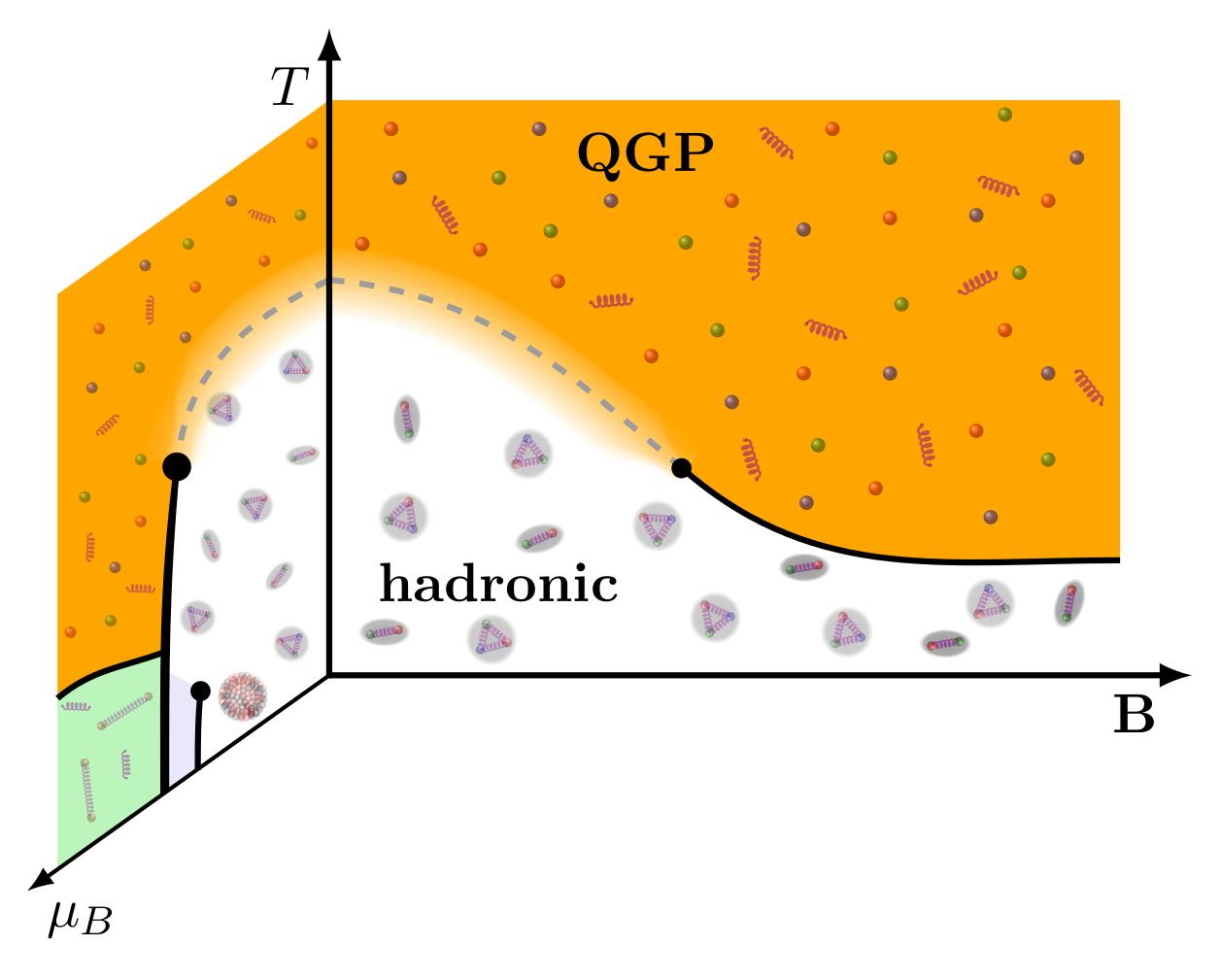}
\caption{Phase diagram extensions in directions of $\mu_I$ (left) and magnetic fields (right). The $\mu_I$ phase diagram features a phase with a BEC of charged pions and a possible BCS phase. The phase diagram at $\mathbf{B}\neq0$ includes a critical endpoint~\cite{DElia:2021yvk}.}
\label{fig-2}       
\end{figure}

After the first pioneering studies~\cite{Kogut:2002tm,Kogut:2002zg}, the continuum phase diagram of isospin asymmetric QCD, $\mu_I\neq0$, has been mapped out at physical quark masses up to $\mu_I\lesssim 325$~MeV~\cite{Brandt:2017oyy,Brandt:2018omg} in the past decade. A sketch of the phase diagram is shown in the left panel of figure~\ref{fig-2}, featuring a phase with a Bose-Einstein condensate (BEC) of charged pions for $\mu_I\geq m_\pi/2$, as predicted by chiral perturbation theory~\cite{Son:2000xc}. The shape of the associated 2nd order phase boundary is almost rectangular and the critical temperature remains below 170~MeV even for $\mu_I\approx 325$~MeV. Perturbation theory predicts the existence of a BCS superconducting phase at large $\mu_I$~\cite{Son:2000xc}, which is actively searched for~\cite{Brandt:2019hel}.
$T_{pc}$ decreases until it hits the 2nd order phase boundary in a pseudo-tricritical point at $(T,\,\mu_I)=(151(7),\,70(5))$~MeV~\cite{Brandt:2017oyy}.
The EoS at vanishing~\cite{Detmold:2012wc,Brandt:2018bwq,Brandt:2022hwy,Abbott:2023coj} and non-zero temperature~\cite{Brandt:2022hwy} has been computed. Novel features are a speed of sound exceeding the conformal value~\cite{Cherman:2009tw} and a negative interaction measure at small temperatures and large $\mu_I$.\footnote{For similar results in a related two-color QCD setting see~\cite{Iida:2022hyy}.} This supports a similar excess in agnostic modelling of the neutron star EoS.

Simulations are also possible for non-zero homogeneous and even spatially modulated~\cite{Brandt:2023dir} $\mathbf{B}$-fields (see~\cite{Endrodi:2024cqn} for a review). The latter are a first step towards a more realistic setting for HICs. For homogeneous external magnetic fields the continuum phase diagram has been determined in Refs.~\cite{Bali:2011qj,Endrodi:2015oba,DElia:2021yvk}, shown schematically in the right panel of figure~\ref{fig-2}. $T_{pc}$ decreases monotonically and the crossover becomes sharper until a critical endpoint is reached~\cite{Endrodi:2015oba}, located between $\vert e\mathbf{B} \vert=4$ and 9~GeV$^2$~\cite{DElia:2021yvk}. The computation of the EoS at $\mathbf{B}\neq0$ is more complex due to flux quantization and the additional breaking of rotational invariance (for details and references see chapter 6 in Ref.~\cite{Endrodi:2024cqn}). Results for the full EoS have been obtained in Ref.~\cite{Bali:2014kia}.

\section{Towards the multi-dimensional parameter space}
\label{sec-2}

To describe physical systems knowledge about the full parameter space is necessary. Currently this can only be achieved using indirect methods starting from the axes discussed above. The two most commonly used methods are Taylor expansion~\cite{Allton:2002zi} and analytic continuation~\cite{deForcrand:2003vyj}. In the former, the pressure is expanded as a Taylor series in $\mu_X/T$ and the expansion coefficients, $\chi_{ijk}^{BQS}$, can be computed at $\mu_X=0$. The latter uses simulations at imaginary $\mu_X$, where the sign problem is absent, and subsequent extrapolations.

\subsection{Results for the equation of state}
\label{sec-2.1}

The current state-of-the-art for the computation of Taylor expansion coefficients at $N_t=8$ are 10th order coefficients~\cite{Adam:2025phc} in a small volume with $LT=2$ and 8th order coefficients with $LT=4$~\cite{Bollweg:2022rps}. Similar results for the coefficients in $\mu_B$-direction have also been extracted from imaginary $\mu_B$ simulations~\cite{DElia:2016jqh,Borsanyi:2018grb}. So far continuum results are only available for the fourth order coefficients, e.g.~\cite{Bollweg:2022rps,Abuali:2025tbd}, and the 6th order coefficient at $LT=2$~\cite{Borsanyi:2023wno}. Note that there is currently a tension between the 6th and 8th order coefficients on coarse lattices, see~\cite{Borsanyi:2023wno}, which might be resolved in the continuum limit. The Taylor coefficients and their temperature dependence give direct access to the EoS. From the newest set of Taylor expansion coefficients at 8th order the EoS has been computed for the $\ev{n_S}=0$ setting and $\mu_Q=0$ in Ref.~\cite{Bollweg:2022fqq} and is estimated to be applicable up to $\mu_B/T\lesssim2.5$~\cite{Bollweg:2022rps}. This agrees with the bound $\mu_B/T\lesssim3.0$ found in~\cite{Borsanyi:2022soo}. At 10th order the range of applicability is expected to be significantly higher. From analytic continuation the most recent result for the equation of state along the $\ev{n_S}=0$ and $ \ev{n_Q}/\ev{n_B}=0.4$ line has been computed in Ref.~\cite{Borsanyi:2022qlh} using an improved extrapolation scheme, the T'-expansion~\cite{Borsanyi:2021sxv}. The T'-expansion leads to an enhanced applicability range up to $\mu_B/T\approx3.5$ or even beyond. The restriction to $\ev{n_S}=0$ is relaxed by the extrapolation of the leading order Taylor coefficient in $\mu_S$-direction to real $\mu_B$. Recently the T'-expansion has also been combined with a 4th order Taylor expansion to obtain the EoS valid for all chemical potential directions~\cite{Abuali:2025tbd}.

To include the effect of $\mathbf{B}$-fields on the finite density EoS, a similar expansion is performed from simulations at $\vert \mathbf{B} \vert \neq0$. Results have been obtained from  a 2nd order Taylor expansion~\cite{Ding:2021cwv,Ding:2023bft,Ding:2025jfz,Ding:2025nyh} and analytic continuation with $\mu_Q=\mu_S=0$~\cite{Astrakhantsev:2024mat} and $\ev{n_S}=0,\, \ev{n_Q}/\ev{n_B}=0.4$~\cite{Borsanyi:2023buy,MarquesValois:2025nzo}. A prominent effect developing at large $\mathbf{B}$ are non-monotonicity of the $\mu_B$ modification of thermodynamic observables with $T$. Furthermore, the Taylor coefficient $\chi_{110}^{BQS}$ shows a particularly strong sensitivity to $\vert \mathbf{B} \vert \neq0$ and has been proposed as a magnetometer for HICs~\cite{Ding:2023bft}. Access to the $\mu_B$ dependence in the BEC phase at $\mu_I\neq0$, can be obtained by a similar Taylor expansion around the simulations at $\mu_I\neq0$~\cite{Brandt:204d2le}.

\subsection{Indirect determination of the phase structure}
\label{sec-2.2}

\begin{figure}[t]
\centering
\includegraphics[width=5.5cm,clip]{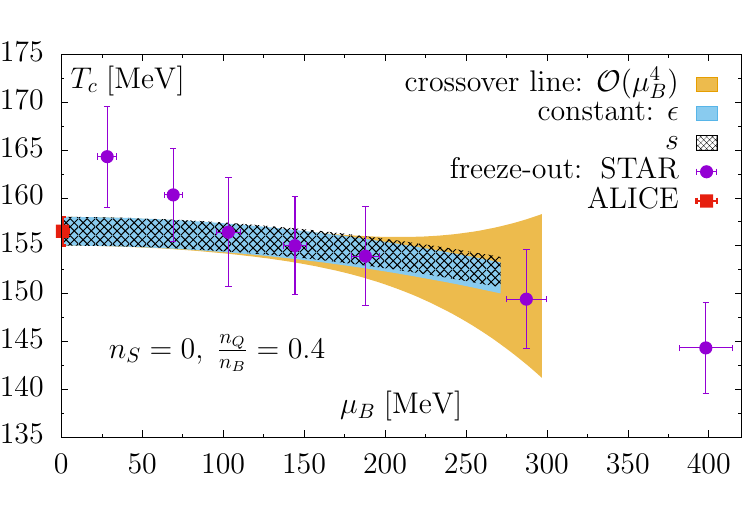} \hspace*{5mm}
\includegraphics[width=5.5cm,clip]{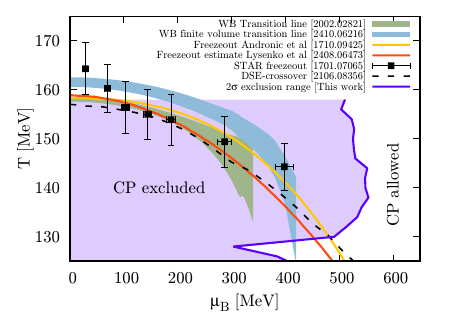}
\caption{{\it Left:} Extension of the chiral crossover to $\mu_B\neq$ at $\ev{n_S}=0$ from Ref.~\cite{HotQCD:2018pds}. {\it Right:} Exclusion region for the CEP location from Ref.~\cite{Borsanyi:2025dyp}.}
\label{fig-3}       
\end{figure}

Indirect methods can also provide information on the phase structure. The extension of the chiral crossover to non-zero $\mu_X$ can be computed from the Taylor expansion of the chiral condensate, for instance. The pseudocritical temperature in $\mu_B$-direction can be expanded as
\begin{equation}
 T_{pc}(\mu_B) = T_{pc}(0) \big[ 1 - \kappa_2 \{\mu_B/T_{pc}(\mu_B)\}^2 - \kappa_4 \{\mu_B/T_{pc}(\mu_B)\}^4 - \ldots \big] \,,
\end{equation}
where the coefficients $\kappa_2$ and $\kappa_4$ depend on the direction in parameter space. Continuum results for $\kappa_2$ for the direction with $\mu_Q=0$ and $\ev{n_S}=0$ or $\mu_s=0$ from Taylor expansion~\cite{Bonati:2018nut,HotQCD:2018pds} or analytic continuation~\cite{Bonati:2015bha,Bellwied:2015rza,Borsanyi:2020fev} agree well and give $\kappa_2\approx0.015$. At the current level of accuracy $\kappa_4$ is consistent with zero~\cite{HotQCD:2018pds,Borsanyi:2020fev}. The crossover extension with the coefficients of Ref.~\cite{HotQCD:2018pds} is shown in the left panel of figure~\ref{fig-3}.

Detecting and locating the elusive CEP at non-zero $\mu_B$ is currently one of the main goals in HIC physics. Its detection using indirect methods is possible in principle, but challenging in practice. Two novel methods for this, which I will briefly discuss here, are (entropy) spinodals~\cite{Shah:2024img} and temperature scaling of Lee-Yang edge (LYE) singularities~\cite{Dimopoulos:2021vrk}. As discussed below, controlling systematic effects is challenging for both methods and future studies will be needed to obtain robust results.

The first method to detect the CEP employs observables which jump at the 1st order phase transition, as the entropy density. Lines of constant observables will cross in the 1st order region and are thus attracted to the CEP, so-called ``critical lensing''~\cite{Nonaka:2004pg}. The CEP can then be located from Taylor expansion or analytic continuation by determining the spinodal and temperature where $(\partial T/\partial s)=(\partial^2 T/\partial s^2)=0$. While a first analysis employing a 2nd order expansion and entropy spinodals found the CEP at $[T^{\rm CEP},\,\mu_B^{\rm CEP}]=[114.3(6.9),\,602(62)]$~MeV~\cite{Shah:2024img}, a subsequent analysis with improved control over systematic effects could only exclude the existence of the CEP up to $\mu_B=450$~MeV~\cite{Borsanyi:2025dyp}, as shown in the right panel of figure~\ref{fig-3}.

The second method utilizes the scaling of real and imaginary part of the (complex) $\mu_B$-location of the LYE singularity (see~\cite{Stephanov:2006dn}), the partition function zero closest to $\mu_B=0$, typically obtained from Pad\'e approximants for the $\mu_B$-dependence of suitable observables. $T^{\rm CEP}$ is then given by the point where the imaginary part of the LYE singularity location vanishes, where the real part determines $\mu_B^{\rm CEP}$. First analyses, using the 8th order Taylor expansion of Ref.~\cite{HotQCD:2018pds} at $N_t=8$~\cite{Basar:2023nkp} or a [4,4] Pad\'e approximation of the first and second order $\mu_B$ cummulants at imaginary $\mu_B$~\cite{Clarke:2024ugt}, respectively, both observed a CEP signal at similar locations. At $[T^{\rm CEP},\,\mu_B^{\rm CEP}]=[105^{+8}_{-18},\,422^{+80}_{-35}]$~MeV for the latter study, for instance. However, the high-accuracy study of Ref.~\cite{Adam:2025phc} highlighted the difficulty to control systematic effects and could only provide an upper bound of $T^{\rm CEP}\lesssim 103$~MeV for the CEP location.

\section{Conclusions}
\label{sec-3}

In the past decades simulations of lattice QCD have lead to predictions for the QCD phase structure and the EoS in several relevant parts of parameter space, in particular where direct simulations are possible, i.e., at non-zero $\mu_I$ and $\mathbf{B}$-fields (discussed in section~\ref{sec-1.2}). For a full description of HICs, however, knowledge about the full parameter space is necessary. Indirect methods are still the state-of-the-art to compute the EoS and to extract information on the phase structure in this region. In the past few years the accuracy and range of applicability of these methods have been increased (see section \ref{sec-2.1}) and new methods to extract information about the possible existence of a CEP have been developed (section \ref{sec-2.2}). The main task for the comming years will be to improve the control over systematic effects and to push the limits further into the chemical potential plane.
\\ \\
{\bf Acknowledgements:} \\
The author is supported by the Deutsche Forschungsgemeinschaft (DFG, German Research Foundation) through CRC-TR 211 -- project number 315477589 -- and MKW NRW under the funding code NW21-024-A. 

%
%

\begin{thebibliography}{}
%
%



\bibitem{Aoki:2006we}
Y.~Aoki \textit{et al.},
Nature \textbf{443}, 675-678 (2006),
\texttt{hep-lat/0611014}
\bibitem{HotQCD:2018pds}
A.~Bazavov \textit{et al.} [HotQCD],
Phys. Lett. B \textbf{795}, 15-21 (2019),
\texttt{1812.08235}
\bibitem{Borsanyi:2020fev}
S.~Borsanyi \textit{et al.}, 
Phys. Rev. Lett. \textbf{125} no.5, 052001 (2020),
\texttt{2002.02821}
\bibitem{Borsanyi:2013bia}
S.~Borsanyi \textit{et al.}, 
Phys. Lett. B \textbf{730}, 99-104 (2014),
\texttt{1309.5258}
\bibitem{HotQCD:2014kol}
A.~Bazavov \textit{et al.} [HotQCD],
Phys. Rev. D \textbf{90}, 094503 (2014),
\texttt{1407.6387}
\bibitem{Bazavov:2017dsy}
A.~Bazavov, P.~Petreczky and J.~H.~Weber,
Phys. Rev. D \textbf{97} no.1, 014510 (2018),
\texttt{1710.05024}
\bibitem{Bresciani:2025vxw}
M.~Bresciani \textit{et al.}, 
Phys. Rev. Lett. \textbf{134} no.20, 201904 (2025),
\texttt{2501.11603}
\bibitem{Kogut:2002tm}
J.~B.~Kogut and D.~K.~Sinclair,
Phys. Rev. D \textbf{66}, 014508 (2002),
\texttt{hep-lat/0201017}
\bibitem{Kogut:2002zg}
J.~B.~Kogut and D.~K.~Sinclair,
Phys. Rev. D \textbf{66}, 034505 (2002),
\texttt{hep-lat/0202028}
\bibitem{Brandt:2017oyy}
B.~B.~Brandt, G.~Endr\H{o}di and S.~Schmalzbauer,
Phys. Rev. D \textbf{97} no.5, 054514 (2018),
\texttt{1712.08190}
\bibitem{Brandt:2018omg}
B.~B.~Brandt and G.~Endr\H{o}di,
Phys. Rev. D \textbf{99} no.1, 014518 (2019),
\texttt{1810.11045}
\bibitem{Son:2000xc}
D.~T.~Son and M.~A.~Stephanov,
Phys. Rev. Lett. \textbf{86}, 592-595 (2001),
\texttt{hep-ph/0005225}
\bibitem{Brandt:2019hel}
B.~B.~Brandt \textit{et al.}, 
Particles \textbf{3} no.1, 80-86 (2020),
\texttt{1912.07451}
\bibitem{Detmold:2012wc}
W.~Detmold, K.~Orginos and Z.~Shi,
Phys. Rev. D \textbf{86}, 054507 (2012),
\texttt{1205.4224}
\bibitem{Brandt:2018bwq}
B.~B.~Brandt \textit{et al.},
Phys. Rev. D \textbf{98} no.9, 094510 (2018),
\texttt{1802.06685}
\bibitem{Brandt:2022hwy}
B.~B.~Brandt, F.~Cuteri and G.~Endr\H{o}di,
JHEP \textbf{07}, 055 (2023),
\texttt{2212.14016}
\bibitem{Abbott:2023coj}
R.~Abbott \textit{et al.} [NPLQCD],
Phys. Rev. D \textbf{108} no.11, 114506 (2023),
\texttt{2307.15014}
\bibitem{Cherman:2009tw}
A.~Cherman, T.~D.~Cohen and A.~Nellore,
Phys. Rev. D \textbf{80}, 066003 (2009),
\texttt{0905.0903}
\bibitem{Iida:2022hyy}
K.~Iida and E.~Itou,
PTEP \textbf{2022} no.11, 111B01 (2022),
\texttt{2207.01253}
\bibitem{Brandt:2023dir}
B.~B.~Brandt \textit{et al.},
JHEP \textbf{11} (2023), 229
\texttt{2305.19029}
\bibitem{Endrodi:2024cqn}
G.~Endr\H{o}di \textit{et al.},
Prog. Part. Nucl. Phys. \textbf{141}, 104153 (2025),
\texttt{2406.19780}
\bibitem{Bali:2011qj}
G.~S.~Bali \textit{et al.},
JHEP \textbf{02}, 044 (2012),
\texttt{1111.4956}
\bibitem{Endrodi:2015oba}
G.~Endr\H{o}di,
JHEP \textbf{07}, 173 (2015),
\texttt{1504.08280}
\bibitem{DElia:2021yvk}
M.~D'Elia \textit{et al.},
Phys. Rev. D \textbf{105} no.3, 034511 (2022),
\texttt{2111.11237}
\bibitem{Bali:2014kia}
G.~S.~Bali \textit{et al.},
JHEP \textbf{08}, 177 (2014),
\texttt{1406.0269}

\bibitem{Allton:2002zi}
C.~R.~Allton \textit{et al.},
Phys. Rev. D \textbf{66}, 074507 (2002)
\texttt{hep-lat/0204010}
\bibitem{deForcrand:2003vyj}
P.~de Forcrand and O.~Philipsen,
Nucl. Phys. B \textbf{673}, 170-186 (2003),
\texttt{hep-lat/0307020}
\bibitem{Adam:2025phc}
A.~Adam \textit{et al.},
\texttt{2507.13254}
\bibitem{Bollweg:2022rps}
D.~Bollweg \textit{et al.} [HotQCD],
Phys. Rev. D \textbf{105} no.7, 074511 (2022),
\texttt{2202.09184}
\bibitem{DElia:2016jqh}
M.~D'Elia, G.~Gagliardi and F.~Sanfilippo,
Phys. Rev. D \textbf{95} no.9, 094503 (2017),
\texttt{1611.08285}
\bibitem{Borsanyi:2018grb}
S.~Borsanyi \textit{et al.},
JHEP \textbf{10}, 205 (2018)
\texttt{1805.04445}
\bibitem{Abuali:2025tbd}
A.~Abuali \textit{et al.},
Phys. Rev. D \textbf{112} no.5, 054502 (2025),
\texttt{2504.01881}
\bibitem{Borsanyi:2023wno}
S.~Borsanyi \textit{et al.},
Phys. Rev. D \textbf{110} no.1, L011501 (2024),
\texttt{2312.07528}
\bibitem{Bollweg:2022fqq}
D.~Bollweg \textit{et al.} [HotQCD],
Phys. Rev. D \textbf{108} no.1, 014510 (2023),
\texttt{2212.09043}
\bibitem{Borsanyi:2022soo}
S.~Borsanyi \textit{et al.},
Phys. Rev. D \textbf{107} no.9, L091503 (2023)
\texttt{2208.05398}
\bibitem{Borsanyi:2022qlh}
S.~Borsanyi \textit{et al.},
Phys. Rev. D \textbf{105} no.11, 114504 (2022)
\texttt{2202.05574}
\bibitem{Borsanyi:2021sxv}
S.~Bors{\'a}nyi \textit{et al.},
Phys. Rev. Lett. \textbf{126} no.23, 232001 (2021),
\texttt{2102.06660}




\bibitem{Ding:2021cwv}
H.~T.~Ding \textit{et al.},
Eur. Phys. J. A \textbf{57} no.6, 202 (2021),
\texttt{2104.06843}
\bibitem{Ding:2023bft}
H.~T.~Ding \textit{et al.},
Phys. Rev. Lett. \textbf{132} no.20, 201903 (2024),
\texttt{2312.08860}
\bibitem{Ding:2025jfz}
H.~T.~Ding \textit{et al.},
Phys. Rev. D \textbf{111} no.11, 114522 (2025),
\texttt{2503.18467}
\bibitem{Ding:2025nyh}
H.~T.~Ding \textit{et al.},
\texttt{2508.07532}
\bibitem{Astrakhantsev:2024mat}
N.~Astrakhantsev \textit{et al.},
Phys. Rev. D \textbf{109} no.9, 094511 (2024),
\texttt{2403.07783}
\bibitem{Borsanyi:2023buy}
S.~Borsanyi \textit{et al.},
PoS \textbf{LATTICE2023}, 164 (2024),
\texttt{2312.15118}
\bibitem{MarquesValois:2025nzo}
A.~D.~Marques Valois \textit{et al.},
PoS \textbf{LATTICE2024}, 177 (2025),
\texttt{2502.01132}
\bibitem{Brandt:204d2le}
B.~B.~Brandt, G.~Endr{\H{o}}di and G.~Mark{\'o},
PoS \textbf{LATTICE2024}, 176 (2025),
\texttt{2411.12918}
\bibitem{Bonati:2018nut}
C.~Bonati \textit{et al.},
Phys. Rev. D \textbf{98} no.5, 054510 (2018),
\texttt{1805.02960}
\bibitem{Bonati:2015bha}
C.~Bonati \textit{et al.},
Phys. Rev. D \textbf{92} no.5, 054503 (2015),
\texttt{1507.03571}
\bibitem{Bellwied:2015rza}
R.~Bellwied \textit{et al.},
Phys. Lett. B \textbf{751}, 559-564 (2015),
\texttt{1507.07510}
\bibitem{Shah:2024img}
H.~Shah \textit{et al.},
\texttt{2410.16206}
\bibitem{Dimopoulos:2021vrk}
P.~Dimopoulos \textit{et al.},
Phys. Rev. D \textbf{105} no.3, 034513 (2022)
\texttt{2110.15933}
\bibitem{Nonaka:2004pg}
C.~Nonaka and M.~Asakawa,
Phys. Rev. C \textbf{71}, 044904 (2005),
\texttt{nucl-th/0410078}
\bibitem{Borsanyi:2025dyp}
S.~Borsanyi \textit{et al.},
\texttt{2502.10267}
\bibitem{Stephanov:2006dn}
M.~A.~Stephanov,
Phys. Rev. D \textbf{73}, 094508 (2006)
\texttt{hep-lat/0603014}
\bibitem{Basar:2023nkp}
G.~Basar,
Phys. Rev. C \textbf{110} no.1, 015203 (2024),
\texttt{2312.06952}
\bibitem{Clarke:2024ugt}
D.~A.~Clarke \textit{et al.},
\texttt{2405.10196}
\end{thebibliography}
%
%

\end{document}